\documentclass[10pt,oneside]{article}
\usepackage{hyperref}
\usepackage{amsmath}
\usepackage{amssymb}
\usepackage{amsfonts}
\usepackage[T1]{fontenc}
\usepackage{relsize,exscale}
\usepackage{ntheorem}
\theoremseparator{:}
\renewcommand{\Bbb}{\mathbb}
\newcommand{\RR}{\Bbb{R}}
\newcommand{\NN}{\Bbb{N}}

\newcommand{\CC}{\Bbb{C}}
\newcommand{\eps}{\varepsilon}
\newtheorem{lemme}{Lemma}[section]
\newtheorem{remarque}{Remark}[section]

\newtheorem{theoreme}{Theorem}[section]
\newtheorem{proposition}{Proposition}[section]

\newtheorem{condition}{Condition}[section]
\evensidemargin=\oddsidemargin
\begin{document}
\title{A proof of the completeness of Lamb modes}
\author{Jean-Luc AKIAN}
\maketitle
\begin{center}

ONERA, Universit\'{e} Paris Saclay F-92322 Ch\^{a}tillon, France

Tel: +33-1-46-73-46-41; Fax: +33-1-46-73-41-43

E-mail address: jean-luc.akian@onera.fr

\end{center}

\maketitle
\begin{abstract}
The aim of this paper is to give a precise proof of the completeness of Lamb modes and associated modes. This proof is relatively simple and short but relies on two powerful mathematical theorems. The first one is a theorem on elliptic systems with a parameter due to Agranovich and Vishik. The second one is a theorem due to Locker which gives a criterion to show the completeness of the set of generalized eigenvectors of a Hilbert-Schmidt discrete operator.
\end{abstract}

{\bf keywords}: 
Lamb modes, completeness, resolvent
\section{Introduction}
Lamb waves are extensively used in nondestructive testing to detect defects in a thin plate because they can scan a wide range of the plate.
The modal formulation of the diffraction of elastic waves by a defect in a plate or of defect detection in a plate by elastic waves relies on the basis property of Lamb modes in the Hilbert space associated to the physical problem (\cite{Bourgeois1}, Assumption 2.6, \cite{Bourgeois2}, Conjecture 2.3, \cite{Bourgeois3}, Conjecture 2).
But this property has not yet been mathematically proved. A weaker property is the completeness of Lamb modes. The basis property and the completeness property of Lamb modes are not equivalent because Lamb modes are solution of a non-self-adjoint spectral problem, thus two Lamb modes corresponding to distinct eigenvalues are not necessarily orthogonal (see section \ref{Compl}).

The paper \cite{Kirrmann} proposes a proof of the completeness of Lamb modes. The proof in this paper relies on a theorem (Theorem 4.2, p.67 of \cite{Kirrmann}) which is not at all proved. In \cite{Kirrmann}, p.68 it is written "The proof of the theorem-but not the theorem itself-is contained in the fundamental paper of Agmon \cite{Agmon}, pp.128-130". But it is not obvious to see a link between Theorem 4.2, p.67 of \cite{Kirrmann} and Theorem 3.2, p.128 of \cite{Agmon}, if that is the case. As is asserted in \cite{Besserer}, p.56 concerning the proof in \cite{Kirrmann}, "However his proof failed in an estimate on circles similar to that presented above. There an unknown inequality had been used that could not be proved". In \cite{Besserer}, pp.54-56 a very short outline of the proof of the completeness of Lamb modes is proposed, but according to \cite{Besserer} the details of the proof are in \cite{Besserer1} (written in German).
As is written in \cite{Pagneux}, p.649  "It is remarkable that, for such a venerable subject, there remain fundamental open questions; e.g., the mathematical proof of the completeness of the Lamb modes has not yet been achieved entirely".

It seems that up to now there is no full proof of the completeness of Lamb modes in the litterature. In the present paper we give a precise, detailed and rigorous proof of the completeness of Lamb modes and associated modes. This proof is relatively simple and short but relies on two powerful mathematical theorems. The first one is a theorem on elliptic systems with a parameter due to Agranovich and Vishik (\cite{Agranovich}) which provides precise estimates in the complex plane of the resolvent of the unbounded operator associated to the physical problem. The second one is a theorem due to Locker (\cite{Locker}) which gives a criterion to show the completeness of the set of generalized eigenvectors of a Hilbert-Schmidt discrete operator ( "a very powerful completeness theorem", \cite{Locker}, p. ix).
The proof in the present paper is provided for traction-free plates on the upper and lower boundary which is the classical case (\cite{Achenbach}, p.220) and is easily extendible to the case of clamped plates on the upper or lower boundary, see Remark \ref{rem1}. It should be noted that in \cite{Kirrmann} and \cite{Besserer} only the case of a plate which is traction-free on one boundary and clamped on the other is considered. 

The paper is organized as follows. In section \ref{Set-up} we estabish the equations of the spectral problem related to Lamb modes and we show that the unbounded operator associated to the physical problem is non-self-adjoint. In section \ref{Res-est}, applying a theorem from \cite{Agranovich} (Theorem \ref{th1}), we give precise resolvent estimates in the complex plane of the unbounded operator associated to the spectral problem for Lamb modes (Theorem \ref{th2}). Finally in section \ref{Compl}, applying a theorem from \cite{Locker} (Theorem \ref{th3}) we prove the completeness of Lamb modes and associated modes (Theorem \ref{th4}).
\section{Set-up of the problem}
\label{Set-up}
\setcounter{equation}{0}
In the sequel we shall use the following notations. The set of natural numbers will be  denoted by $\NN$ (containing $0$) and the set of positive natural number by $\NN^*$ (= $\NN \setminus \{0\}$). If $n, m \in \NN^*$ and $\Omega$ is an open set of $\RR^n$, the set of ${\cal C}^{\infty}$ functions from $\Omega$ with values in $\RR^m$ and with compact support in $\Omega$ will be denoted by ${\cal C}^{\infty}_0(\Omega, \RR^m)$, with similar notations for functions with values in $\CC^m$. If $k \in \NN^*$, the set of functions from $\Omega$ with values in  $\RR^m$ whose components are in the Sobolev space $H^k(\Omega)$ will be denoted by $H^k(\Omega,\RR^m)$, with similar notations for functions with values in $\CC^m$.
%
The inner product in $H^k(\Omega,\RR^m)$ or $H^k(\Omega,\CC^m)$ will be denoted by $(.,.)_{m,\Omega}$, the associated norm by $||.||_{m,\Omega}$ and the associated semi-norm by $|.|_{m,\Omega}$.
The identity of a vector space will be denoted by $I$ regardless of the vector space. Recall that the word "iff" means "if and only if".

Let us consider a linearly elastic plate of thickness $2h$ occupying the open set $\Omega$ of $\RR^3$:
\begin{equation}
\label{eq1a}
\Omega = \{x \in \RR^3, \, -h < x_1 < h\}.
\end{equation}
The plate is assumed to be homogeneous and isotropic (with Lam\'{e} coefficients $\lambda$ and $\mu$ such that $3 \lambda + 2 \mu >0$ and $\mu >0$, see \cite{Salencon}, p.341), with mass density $\rho$ and traction-free on the upper and lower boundary. Denote by $u=(u_i)$, $\eps_{ij}(u)$, $\sigma_{ij}(u)$ the displacement field, the components of the strain tensor and the components of the stress tensor associated to $u$. The elastodynamics equations and boundary conditions for the plate are written as follows (the derivative with respect to time and to $x_i$ being denoted by a dot and by $\partial_i$):
\begin{equation}
\label{eq2}
\partial_j\sigma_{ij}(u) = \rho \ddot{u_i} \mbox{  in } \Omega,
\end{equation}
with
\begin{equation}
\label{eq4}
\sigma_{ij}(u) = \lambda (div u) \delta_{ij} + 2 \mu \eps_{ij}(u)\mbox{  in } \Omega,
\end{equation}
%
\begin{equation}
\label{eq4a}
\eps_{ij}(u) = \frac{1}{2}\left(\partial_{j} u_i + \partial_{i} u_j \right)\mbox{  in } \Omega
\end{equation}
and 
\begin{equation}
\label{eq4b}
\sigma_{i1}(u)(x_1 = \pm h) = 0.
\end{equation}
We seek the displacement field $u$ solution of the elastodynamics equations
\eqref{eq2}, \eqref{eq4}, \eqref{eq4a}, \eqref{eq4b} under the form of harmonic waves propagating in the $x_3$ direction and independant of $x_2$ because of the invariance of the physical properties in the $x_2$ direction.
We use the following notations: 
\begin{equation}
\label{eq6a}
u(x_1,x_3,t) = \mbox{ Re}(\tilde{u}(x_1,x_3,t)),
\end{equation}
\begin{equation}
\label{eq6c}
\tilde{u}(x_1,x_3,t) = v(x_1) e^{i(\beta x_3 - \omega t)}, \,v = \left (
\begin{array}{c}
v_1 \\ v_2 \\ v_3
\end{array}
\right ),
\end{equation}
where $\omega >0$ is fixed and $\beta \in \CC$ is to be determined.
Since $\mbox{ Re}(\tilde{u}(x_1,x_3,t+\pi/2\omega))$ = $\mbox{ Re}(-i\tilde{u}(x_1,x_3,t))$ = $\mbox{ Im}(\tilde{u}(x_1,x_3,t))$
and
$\mbox{ Re}(\ddot{\tilde{u}}(x_1,x_3,t+\pi/2\omega))$ = $\mbox{ Re}(-i\ddot{\tilde{u}}(x_1,x_3,t))$ = $\mbox{ Im}(\ddot{\tilde{u}}(x_1,x_3,t))$, in equations \eqref{eq2}, \eqref{eq4}, \eqref{eq4a}, \eqref{eq4b} $u$ may be replaced by $\tilde{u}$.

The displacement field $\tilde{u}$ (with values in $\CC^3$) is formally characterized by the following variational problem: for all "regular" displacement field $\delta u$ with values in $\CC^3$ and with compact support in $\overline{\Omega}$, 
\begin{equation}
\label{eq1}
m(\ddot{\tilde{u}},\delta u) + k(\tilde{u},\delta u) =0
\end{equation}
with the notations
\begin{equation}
\label{eq2a}
k(u, \delta u) = \int _{\Omega} \sigma_{ij}(u) \eps_{ij}(\overline{\delta u}) 
\end{equation}
and
\begin{equation}
\label{eq3}
m(u, \delta u) = \int _{\Omega} \rho u_i \overline{\delta u_i}.
\end{equation}
We have the formulas:
\begin{equation}
\label{eq5}
\sigma_{ij}(\tilde{u}) \eps_{ij}(\overline{\delta u}) = \lambda (div \tilde{u}) (div \overline{\delta u}) + 2 \mu \eps_{ij}(\tilde{u}) \eps_{ij}(\overline{\delta u}),
\end{equation}
\begin{equation}
\label{eq7}
{\mbox {div} } \tilde{u} = 
(\partial_1 v_1 + i \beta v_3)
e^ {i(\beta x_3 - \omega t)},
\end{equation}
\begin{equation}
\label{eq8}
\eps_{11} (\tilde{u}) = \partial_1 v_1  e^ {i(\beta x_3 - \omega t)},\, \eps_{22} (\tilde{u}) =  0, \, \eps_{33} (\tilde{u}) = i\beta v_3  e^ {i(\beta x_3 - \omega t)},
\end{equation}
\begin{equation}
\label{eq9}
\eps_{12} (\tilde{u}) = 1/2 \partial_1 v_2  e^ {i(\beta x_3 - \omega t)},\, \eps_{13} (\tilde{u}) = 1/2 (\partial_1 v_3 + i \beta v_1) e^ {i(\beta x_3 - \omega t)}, \, \eps_{23} (\tilde{u}) = 1/2 i\beta v_2  e^ {i(\beta x_3 - \omega t)}.
\end{equation}
We choose $\delta u$ under the form
\begin{equation}
\label{eq6}
\delta u(x_1,x_3) = 
\delta v(x_1)
\varphi(x_3),\,
\delta v = \left (
\begin{array}{c}
\delta v_1 \\ \delta v_2 \\ \delta v_3
\end{array}
\right ),
\end{equation}
where
$\varphi \in {\cal C}^{\infty}_0(\RR, \CC)$. The following formulas hold: 
\begin{equation}
\label{eq7a}
{\mbox {div} } \delta u = 
\partial_1 \delta v_1 \varphi + \delta v_3 \partial_3\varphi,
\end{equation}
\begin{equation}
\label{eq8a}
\eps_{11} (\delta u) = \partial_1 \delta v_1 \varphi,\, \eps_{22} (\delta u) =  0, \, \eps_{33} (\delta u) = \delta v_3 \partial_3\varphi,
\end{equation}
\begin{equation}
\label{eq9a}
\eps_{12} (\delta u) = 1/2 \partial_1 \delta v_2  \varphi,\, \eps_{13} (\delta u) = 1/2 (\partial_1 \delta v_3 \varphi + \delta v_1 \partial_3\varphi), \, \eps_{23} (\delta u) = 1/2  \delta v_2  \partial_3 \varphi.
\end{equation}
We get:
\begin{equation}
\label{eq10}
{\mbox {div} } \tilde{u} {\mbox {div} } \overline{\delta u} = 
(\partial_1 v_1 + i \beta v_3)
(\partial_1 \overline{\delta v_1} \overline{\varphi} + \overline{\delta v_3} \overline{\partial_3\varphi})
e^ {i(\beta x_3 - \omega t)},
\end{equation}
\begin{eqnarray}
\label{eq11}
&&\eps_{ij}(\tilde{u}) \eps_{ij}(\overline{\delta u}) = \{\partial_1 v_1 \partial_1 \overline{\delta v_1}\overline{\varphi} + i \beta v_3  \overline{\delta v_3}\overline{\partial_3\varphi}+ \nonumber \\ 
&&+1/2 \partial_1 v_2 \partial_1 \overline{\delta v_2} \overline{\varphi} +1/2 (\partial_1 v_3  + i \beta v_1) (\partial_1 \overline{\delta v_3} \overline{\varphi} + \overline{\delta v_1}\overline{\partial_3\varphi}) + \nonumber \\
&&1/2 i \beta v_2 \overline{\delta v_2} \overline{\partial_3\varphi}\}
e^ {i(\beta x_3 - \omega t)},
\end{eqnarray}
\begin{equation}
\label{eq12}
\rho \ddot{\tilde{u}}_i \overline{\delta u}_i = -\rho \omega^2 (v_1 \overline{\delta v_1}+ v_2 \overline{\delta v_2}+ v_3 \overline{\delta v_3})\overline{\varphi} e^ {i(\beta x_3 - \omega t)}.
\end{equation}

Choose $\varphi$ under the form $\varphi(x_3)= e^{i \overline{\beta} x_3} \psi(x_3)$ where $\psi  \in {\cal C}^{\infty}_0(\RR, \RR)$ and $\int_{\RR} \psi =1$. Then $\overline{\varphi(x_3)}  e^ {i\beta x_3} =\psi(x_3)$ and $\overline{\partial_3\varphi(x_3)}  e^ {i\beta x_3} =\partial_3\psi(x_3) - i \beta \psi(x_3)$.
Taking into account that $\int_{\RR} \partial_3\psi =0$ and gathering all the previous results we obtain a mathematical formulation of the problem: with the notation $\omega_h = (-h,h)$, for a fixed $\omega >0$, find $\beta \in \CC$ such that $\exists$ $v$ = $(v_1,v_2,v_3)^T$ $\neq 0$ $\in H^1(\omega_h, \CC^3)$, such that for all $\delta v$ = $(\delta v_1,\delta v_2,\delta v_3)^T$ $\in H^1(\omega_h, \CC^3)$, 
\begin{equation}
\label{eq13}
a(v, \delta v) + \beta b(v, \delta v) + \beta^2 c(v, \delta v)=0
\end{equation}
where for all $v$, $\delta v$ $\in H^1(\omega_h, \CC^3)$,
\begin{equation}
\label{eq13a}
a(v, \delta v) = a_0(v, \delta v) - \omega^2 l(v, \delta v),
\end{equation}
\begin{equation}
\label{eq14}
a_0(v, \delta v) = \int_{\omega_h} (\lambda + 2 \mu) \partial_1 v_1 \partial_1\overline{\delta v_1} + \mu(\partial_1 v_2 \partial_1\overline{\delta v_2} + \partial_1 v_3 \partial_1\overline{\delta v_3}),
\end{equation}
\begin{equation}
\label{eq14a}
l(v, \delta v) = \int_{\omega_h} \rho (v_1 \overline{\delta v_1}+ v_2 \overline{\delta v_2}+ v_3 \overline{\delta v_3}),
\end{equation}
\begin{equation}
\label{eq15}
b(v, \delta v) = \int_{\omega_h} \lambda (-i \partial_1 v_1 \overline{\delta v_3} + i v_3  \partial_1 \overline {\delta v_1}) + \mu (-i \partial_1 v_3 \overline{\delta v_1} + i v_1 \partial_1 \overline{\delta v_3}),
\end{equation}
\begin{equation}
\label{eq16}
c(v, \delta v) = \int_{\omega_h} (\lambda + 2 \mu)  v_3 \overline{\delta v_3} +  \mu (v_1 \overline{\delta v_1} + v_2 \overline{\delta v_2}).
\end{equation}
Then $v$ is solution of \eqref{eq13} iff $v \in H^2(\omega_h,\CC^3)$ and $v$ satisfies the following equations
\begin{equation}
\label{eq16b}
(\lambda+ 2 \mu) \partial_{11} v_1+ (\lambda + \mu) i \beta \partial_1 v_3 = (\mu \beta^2 - \omega^2 \rho) v_1 \mbox{ in } \omega_h,
\end{equation}
\begin{equation}
\label{eq16c}
\mu \partial_{11} v_2 = (\mu \beta^2 - \omega^2 \rho) v_2  \mbox{ in } \omega_h,
\end{equation}
\begin{equation}
\label{eq16d}
\mu \partial_{11} v_3 + (\lambda + \mu) i \beta \partial_1 v_1 = ((\lambda + 2\mu) \beta^2 - \omega^2 \rho) v_3  \mbox{ in } \omega_h,
\end{equation}
and boundary conditions
\begin{equation}
\label{eq16e}
(\lambda+ 2 \mu) \partial_1 v_1(\pm h) + \lambda i \beta  v_3( \pm h)= 0, 
\end{equation}
\begin{equation}
\label{eq16f}
\mu \partial_1 v_2(\pm h) = 0,
\end{equation}
\begin{equation}
\label{eq16g}
\mu (\partial_1 v_3 ( \pm h)+  i \beta v_1 (\pm h))= 0.
\end{equation}
The variational problem \eqref{eq13} is splitted in two independant problems: one for the components $v_1, v_3$ (Lamb modes) and one for the component $v_2$ ($SH$ modes).
In the sequel we shall examine the spectral problem for Lamb modes.
Lamb modes are solution of the following problem: for a fixed $\omega>0$, find $\beta \in \CC$ ($i \beta$ will be called a Lamb eigenvalue) such that $\exists$ $v_L$ = $(v_1,v_3)^T$ $\neq 0$ $\in H^1(\omega_h, \CC^2)$ ($v_L$ will be called a Lamb mode), such that for all $\delta v_L$ = $(\delta v_1,\delta v_3)^T$ $\in H^1(\omega_h, \CC^2)$,
\begin{equation}
\label{eq13b}
a_L(v_L, \delta v_L) + \beta b_L(v_L, \delta v_L) + \beta^2 c_L(v_L, \delta v_L)=0
\end{equation}
where for all $v_L$, $\delta v_L$ $\in H^1(\omega_h, \CC^2)$,
\begin{equation}
\label{eq13c}
a_L(v_L, \delta v_L) = a_{0,L}(v_L, \delta v_L) - \omega^2 l_L(v_L, \delta v_L),
\end{equation}
\begin{equation}
\label{eq14b}
a_{0,L}(v_L, \delta v_L) = \int_{\omega_h} (\lambda + 2 \mu) \partial_1 v_1 \partial_1\overline{\delta v_1}  + \mu \partial_1 v_3 \partial_1\overline{\delta v_3},
\end{equation}
\begin{equation}
\label{eq14c}
l_L(v_L, \delta v_L) = \int_{\omega_h} \rho (v_1 \overline{\delta v_1}+ v_3 \overline{\delta v_3}),
\end{equation}
\begin{equation}
\label{eq15a}
b_L(v_L, \delta v_L) = \int_{\omega_h} \lambda (-i \partial_1 v_1 \overline{\delta v_3} + i v_3  \partial_1 \overline {\delta v_1}) + \mu (-i \partial_1 v_3 \overline{\delta v_1} + i v_1 \partial_1 \overline{\delta v_3}),
\end{equation}
\begin{equation}
\label{eq16a}
c_L(v_L, \delta v_L) = \int_{\omega_h} (\lambda + 2 \mu)  v_3 \overline{\delta v_3} +  \mu v_1 \overline{\delta v_1}.
\end{equation}

With the notations
\begin{equation}
\label{eq16h}
A = \left(
\begin{array}{cc}
\lambda+2\mu & 0 \\
0 & \mu
\end{array}
\right), \,
B = \left(
\begin{array}{cc}
0 & \lambda+\mu\\
\lambda + \mu & 0
\end{array}
\right), \,
C = \left(
\begin{array}{cc}
\mu & 0 \\
0 & \lambda + 2\mu
\end{array}
\right)
\end{equation}
and
\begin{equation}
\label{eq16i}
D = \left(
\begin{array}{cc}
0 & \lambda\\
\mu & 0
\end{array}
\right),
\end{equation}
for all sufficiently regular $v_L$, $\delta v_L$, one can write
\begin{eqnarray}
\label{eq16j}
a_{0,L}(v_L,\delta v_L) &=& (A \partial_{1}v_L, \partial_1\delta v_L)_{0,\omega_h}  \nonumber \\
 &=& -(A \partial_{11}v_L,\delta v_L)_{0,\omega_h} + [A \partial_1 v_L \cdot \delta v_L]_{-h}^{h} \nonumber \\
&=&-(A v_L, \partial_{11} \delta v_L)_{0,\omega_h} + [A v_L \cdot \partial_1\delta v_L]_{-h}^{h},
\end{eqnarray}
\begin{eqnarray}
\label{eq16k}
b_{L}(v_L,\delta v_L)&= &-i (B \partial_{1}v_L,\delta v_L)_{0,\omega_h} + i[D v_L \cdot \delta v_L]_{-h}^{h}\nonumber \\
&= &i(B v_L,\partial_{1}\delta v_L)_{0,\omega_h} - i[v_L \cdot D\delta v_L]_{-h}^{h}
\end{eqnarray}
and
\begin{equation}
\label{eq16l}
c_{L}(v_L,\delta v_L)= (C v_L,\delta v_L)_{0,\omega_h}.
\end{equation}
The characterization \eqref{eq13b} of Lamb modes is equivalent to $v_L \in H^2(\omega_h,\CC^2)$ and to equations \eqref{eq16b}, \eqref{eq16d}, \eqref{eq16e} and \eqref{eq16g} which read as follows:
\begin{equation}
\label{eq17}
A \partial_{11}v_L+ i \beta B \partial_1 v_L + (\omega^2 \rho-\beta^2 C) v_L = 0\mbox{ in } \omega_h
\end{equation}
and 
\begin{equation}
\label{eq18}
A \partial_1 v_L(\pm h) + i \beta D v_L(\pm h) = 0. 
\end{equation}
With the notations
\begin{equation}
\label{eq19}
V_L = \left(
\begin{array}{c}
v_L\\
i \beta v_L
\end{array}
\right),
\end{equation}
\begin{equation}
\label{eq20}
\tilde{\cal A}= \left(
\begin{array}{cc}
0 & I\\
- C^{-1}(\omega^2 \rho + A \partial_{11}) & -C^{-1} B \partial_1
\end{array}
\right)
\end{equation}
and
\begin{equation}
\label{eq21}
\tilde{\cal B}=
(
\begin{array}{cc}
A \partial_1& D
\end{array}
),
\end{equation}
we obtain
\begin{equation}
\label{eq22}
i \beta V_L = \tilde{\cal A} V_L
\end{equation}
and
\begin{equation}
\label{eq23}
\tilde{\cal B} V_L (\pm h) = 0.
\end{equation}
If $H$ is a Hilbert space, an unbounded operator $T$ : $H \rightarrow H$ is a linear operator $T$ from a linear subspace $D(T) \subset H$ (the domain of $T$) into $H$. An unbounded operator $T$ is closed iff by definition its graph $G(T) =\{(x,Tx), x \in D(T) \}$ is closed in $H \times H$. The graph norm of an element $(x,Tx)$ of $G(T)$ is by definition $||x|| + ||Tx||$.
With the notations
\begin{equation}
\label{eq24}
W= H^2(\omega_h, \CC^2),\, V= H^1(\omega_h, \CC^2), \, H = L^2(\omega_h, \CC^2),
\end{equation}
\begin{equation}
\label{eq25}
{\cal H} = V \times H,\, {\cal V} = W \times V,
\end{equation}
${\tilde{\cal A}}$ and ${\tilde{\cal B}}$ define an unbounded operator ${\cal A}$ in ${\cal H}$ with domain
\begin{equation}
\label{eq26}
D({\cal A}) = \{U \in {\cal H},\, \tilde{\cal A} U \in {\cal H},\, \tilde{\cal B} U (\pm h) = 0\}=
\{U \in {\cal V},\, \tilde{\cal B} U (\pm h) = 0\}
\end{equation}
and by
\begin{equation}
\label{eq26a}
\forall U \in D({\cal A}), \,{\cal A}U = \tilde{\cal A}U.
\end{equation}
The following proposition holds:
\begin{proposition}
\label{Prop1}
The unbounded operator ${\cal A}$ is closed and its domain $D({\cal A})$ is dense in ${\cal H}$.
\end{proposition}
{\bf Proof}
On $D({\cal A})$ the graph norm and the norm on ${\cal V}$ are equivalent then the operator ${\cal A}$ is closed.
On the other hand, suppose that $U = (U_1,U_2) \in {\cal H}$ is orthogonal to $D({\cal A})$. Then it is orthogonal to $\{0\} \times {\cal C}^{\infty}_0(\omega_h, \CC^2)$ that is to say $U_2$ is orthogonal to ${\cal C}^{\infty}_0(\omega_h, \CC^2)$ for the scalar product of $H$, so that $U_2=0$. If $V_1 \in W$, it is possible to construct $V_2 \in V$ such that $(V_1,V_2) \in D({\cal A})$. Then for all $V_1$ $\in W$, $U_1$ is orthogonal to $V_1$ for the scalar product of $V$. Since $W$ is dense in $V$ one obtains $U_1=0$ and the conclusion follows. $\Box$

Let us equip the Hilbert space ${\cal H}$ with the scalar product (equivalent to the natural scalar product on ${\cal H}$) : for all $U = (U_1,U_2)$ and $\delta U = (\delta U_1, \delta U_2)$ $\in$ ${\cal H}$,
\begin{equation}
\label{eq26b}
(U,\delta U)_{\cal H}= a_{0,L}(U_1,\delta U_1) + (U_1,\delta U_1)_{0,\omega_h} + c(U_2,\delta U_2).
\end{equation}
By definition the adjoint ${\cal A}^*$ of ${\cal A}$ (which is well-defined since the domain of ${\cal A}$ is dense in ${\cal H}$) is an unbounded operator with domain:
\begin{equation}
\label{eq26d}
D({\cal A}^*)= \{ \delta U \in {\cal H}, \, \exists C> 0 \mbox{ such that } \forall U \in D({\cal A}), |({\cal A} U,\delta U)_{\cal H}| \leq C ||U||_{\cal H}\}.
\end{equation}
For all $U \in D({\cal A})$, for all $\delta U$ $\in {\cal H}$, 
\begin{eqnarray}
\label{eq26c}
&&({\cal A} U,\delta U)_{\cal H}= a_{0,L}(U_2,\delta U_1)
 + (U_2,\delta U_1)_{0,\omega_h} \nonumber \\
&&-\omega^2 l_L(U_1, \delta U_2)
-(A \partial_{11}U_1, \delta U_2)_{0,\omega_h}
-(B \partial_{1}U_2,\delta U_2)_{0,\omega_h}.
\end{eqnarray}
Taking $U \in {\cal C}^{\infty}_0(\omega_h, \CC^2) \times {\cal C}^{\infty}_0(\omega_h, \CC^2) \subset D({\cal A})$ in the characterization of $D({\cal A}^*)$, one obtains $\delta U \in {\cal V}$.
From \eqref{eq16j}, for all $U \in D({\cal A})$, for all $\delta U$ $\in {\cal V}$, 
\begin{eqnarray}
\label{eq26e}
({\cal A} U,\delta U)_{\cal H}&=& 
-(A U_2,\partial_{11} \delta U_1)_{0,\omega_h} + [A U_2 \cdot \partial_1\delta U_1]_{-h}^{h}
+ (U_2,\delta U_1)_{0,\omega_h}
-\omega^2 l_L(U_1, \delta U_2)\nonumber \\
&&
+ a_{0,L}(U_1, \delta U_2)
-[A\partial_1 U_1 \cdot \delta U_2]_{-h}^h
+(B U_2,\partial_{1}\delta U_2)_{0,\omega_h} 
-[B U_2 \cdot \delta U_2]_{-h}^h.
\end{eqnarray}
Since $U \in D({\cal A})$ the boundary part in \eqref{eq26e} is
\begin{eqnarray}
\label{eq26f}
&&[A U_2 \cdot \partial_1\delta U_1]_{-h}^{h}
-[A\partial_1 U_1 \cdot \delta U_2]_{-h}^h
-[B U_2 \cdot \delta U_2]_{-h}^h \nonumber \\
&&= [U_2 \cdot (A\partial_1\delta U_1-D\delta U_2)]_{-h}^{h}.
\end{eqnarray}
Consequently with the notation
\begin{equation}
\label{eq26g}
\tilde{\cal B}^*=
(
\begin{array}{cc}
A \partial_1& -D
\end{array}
),
\end{equation}
the domain of ${\cal A}^*$ is 
\begin{equation}
\label{eq26h}
D({\cal A}^*) = 
\{U \in {\cal V},\, \tilde{\cal B}^* U (\pm h) = 0\}.
\end{equation}
Since $D({\cal A}) \neq D({\cal A}^*)$ it ensues that ${\cal A}$ is non-self-adjoint.
For all $U \in D({\cal A})$, for all $\delta U \in D({\cal A}^*)$,
\begin{eqnarray}
\label{eq26i}
({\cal A} U,\delta U)_{\cal H}&=& 
-(U_2,A \partial_{11} \delta U_1)_{0,\omega_h} 
+ (U_2,\delta U_1)_{0,\omega_h}
-\omega^2 l_L(U_1, \delta U_2)\nonumber \\
&&
+ a_{0,L}(U_1, \delta U_2)
+(U_2,B \partial_{1}\delta U_2)_{0,\omega_h}.
\end{eqnarray}
For all $\delta U_2 \in V$, the map $u \in V$ $\rightarrow$ $a_{0,L}(u, \delta U_2)-\omega^2 l_L(u, \delta U_2)$ is continuous on $V$ equipped with the scalar product $a_{0,L}(.,.) + (.,.)_{0,\omega_h}$. By the Riesz representation theorem there exists a unique $R(\delta U_2) \in V$ such that for all $u \in V$,

\begin{equation}
\label{eq26j}
a_{0,L}(u, \delta U_2)-\omega^2 l_L(u, \delta U_2) = 
a_{0,L}(u,R(\delta U_2)) + (u,R(\delta U_2))_{0,\omega_h}.
\end{equation}
In view of \eqref{eq26i} and \eqref{eq26j} we obtain: 
%
\begin{equation}
\label{eq26k}
\forall U \in D({\cal A}^*), \,{\cal A}^*U = \tilde{\cal A}^*U,
\end{equation}
where
\begin{equation}
\label{eq26l}
\tilde{\cal A}^*= \left(
\begin{array}{cc}
0 & R\\
 C^{-1}(-A \partial_{11}+I) & C^{-1} B \partial_1
\end{array}
\right).
\end{equation}
\section{Resolvent estimates}
\label{Res-est}
\setcounter{equation}{0}
Before proceeding we first set  some definitions and results about Hilbert space operators (\cite{Locker}, p.21). 
If $H$ is a Hilbert space and $T$ is an unbounded closed linear operator in $H$, the resolvent set of $T$ denoted by $\rho(T)$ is the set of $\lambda \in \CC$ such that the operator $T - \lambda I$ is a one-to-one mapping from its domain $D(T-\lambda I)$ = $D(T)$ onto the Hilbert space $H$ (in that case $(T- \lambda I)^{-1}$ is a bounded operator in $H$). The spectrum of $T$ is the complement of $\rho(T)$ in $\CC$: $\sigma(T) = \CC \setminus \rho(T)$. If $\lambda \in \rho(T)$, the operator $(T - \lambda I)^{-1}$ is called the resolvent of $T$.

Let us now study the resolvent of ${\cal A}$. If $\beta \in \CC$ and $F= (F_1,F_2)\in {\cal H}$, let us seek the solutions $U \in D({\cal A})$ of the equation:
\begin{equation}
\label{eq27a}
({\cal A} - i \beta I)U =F.
\end{equation}
This equation is equivalent to
\begin{equation}
\label{eq27}
U_2= i \beta U_1+ F_1\mbox{ in } \omega_h
\end{equation}
and
\begin{equation}
\label{eq28}
A \partial_{11} U_1 +i\beta B \partial_1 U_1 +(\omega^2\rho - \beta^2 C) U_1= -i \beta C F_1 - B \partial_1 F_1 - C F_2 \mbox{ in } \omega_h.
\end{equation}
The condition $ U  \in D({\cal A})$ is equivalent to 
\begin{equation}
\label{eq27b}
A \partial_1 U_1 (\pm h) + i \beta D U_1(\pm h) = -D  F_1(\pm h).
\end{equation}
If $U =(U_1,U_2) \in D({\cal A})$ satisfies \eqref{eq27}, \eqref{eq28} then $v_L =U_1$ satisfies the following variational formulation: for all $\delta v_L$ = $(\delta v_1,\delta v_3)^T$ $\in H^1(\omega_h, \CC^2)$,
\begin{eqnarray}
\label{eq29}
&&a_L(v_L, \delta v_L) + \beta b_L(v_L, \delta v_L) + \beta^2 c_L(v_L, \delta v_L)= \nonumber \\
&&(CF_2,\delta v_L)_{0,\omega_h} + (B\partial_1 F_1,\delta v_L)_{0,\omega_h} +i \beta (CF_1,\delta v_L)_{0,\omega_h} - [ DF_1\cdot \delta v_L ]_{-h}^{h}.
\end{eqnarray}
Conversely if $v_L$ satisfies the variational formulation \eqref{eq29} then $U=(U_1,U_2)$ where $U_1= v_L$ and $U_2$ is given by \eqref{eq27} is such that $U \in D({\cal A})$ and $U$ satisfies \eqref{eq28}.
But with the notation $\Omega_h = \omega_h \times (0,1)$, for all $v \in H^1(\omega_h,\CC^3)$, for all $\beta \in \RR$, 
\begin{equation}
\label{eq30}
a(v,v) + \beta b(v,v) + \beta^2 c(v,v)= \int_{\Omega_h} (\sigma_{ij}(u)\eps_{ij}(\overline{u}) - \omega^2 \rho u_i \overline{u_i})
\end{equation}
where 
\begin{equation}
\label{eq30a}
u(x_1,x_3) = v(x_1) e^{i\beta x_3}.
\end{equation}
Owing to Korn inequality, there exist two constants $C_1,\, C_2 >0$ such that for all $u \in H^1(\Omega_h,\CC^3)$,
\begin{equation}
\label{eq31}
\int_{\Omega_h} \sigma_{ij}(u)\eps_{ij}(\overline{u}) \geq
C_1 ||u||_{1,\Omega_h}^2 -C_2 ||u||_{0,\Omega_h}^2.
\end{equation}
But if $u$ is given by \eqref{eq30a} then
\begin{equation}
\label{eq32}
|u|_{1,\Omega_h}^2 = |v|_{1,\omega_h}^2 + \beta^2 ||v||_{0,\omega_h}^2
\end{equation}
and
\begin{equation}
\label{eq33}
||u||_{0,\Omega_h}^2 =||v||_{0,\omega_h}^2.
\end{equation}
From \eqref{eq30}, \eqref{eq31}, \eqref{eq32} and \eqref{eq33}, there exist $C > 0$ and $\beta_0 >0$ such that for all $\beta \in \RR$, $|\beta| \geq \beta_0$, for all $v \in H^1(\omega_h,\CC^3)$,
\begin{equation}
\label{eq35}
a(v,v) + \beta b(v,v) + \beta^2 c(v,v) \geq C (|v|_{1,\omega_h}^2 + \beta^2 ||v||_{0,\omega_h}^2) \geq C ||v||_{1,\omega_h}^2 .
\end{equation}
Consequently there exist $C > 0$ and $\beta_0 >0$ such that for all $\beta \in \RR$, $|\beta| \geq \beta_0$, for all $v_L \in H^1(\omega_h,\CC^2)$,
\begin{equation}
\label{eq36}
a_L(v_L,v_L) + \beta b_L(v_L,v_L) + \beta^2 c_L(v_L,v_L) \geq C (|v_L|_{1,\omega_h}^2 + \beta^2 ||v_L||_{0,\omega_h}^2) \geq C ||v_L||_{1,\omega_h}^2.
\end{equation}
In view of  \eqref{eq36}, \eqref{eq27} and Lax-Milgram theorem, if $\beta \in \RR$, $|\beta| \geq \beta_0$, then ${\cal A} - i \beta I$ is one-to-one from $D({\cal A})$ onto ${\cal H}$ and $({\cal A} - i \beta I)^{-1}$ is continuous from ${\cal H}$ into ${\cal H}$ which implies $i\beta \in \rho({\cal A})$.

Taking \eqref{eq27} and \eqref{eq28} into account, if $\beta \in \RR$, $|\beta| \geq \beta_0$, then $({\cal A} - i \beta I)^{-1}$ is continuous from ${\cal H}$ into ${\cal V}$. Since the embedding from ${\cal V}$ into ${\cal H}$ is compact it can be inferred  and $({\cal A} - i \beta I)^{-1}$ is a compact operator in ${\cal H}$ and thus ${\cal A}$ has a compact resolvent.

Let us now examine the behavior of the resolvent for $\beta \in \CC$. If $\beta \in \CC$, $\beta = a+ i b$, then for all $v_L \in H^1(\omega_h,\CC^2)$,
\begin{equation}
\label{eq103a}
{\mbox {Re}}\left\{a_L(v_L,v_L) + \beta b_L(v_L,v_L) + \beta^2 c_L(v_L,v_L)\right\}  = 
a_L(v_L,v_L) + a b_L(v_L,v_L) + (a^2 - b^2) c_L(v_L,v_L).
\end{equation}
Let us choose $0 < \alpha <1$. Due to \eqref{eq36} it follows that if $|a| \geq \beta_0$ and $|b| \leq \alpha |a|$, then for all $v_L \in H^1(\omega_h,\CC^2)$,
\begin{eqnarray}
\label{eq103b}
&&a_L(v_L,v_L) + a b_L(v_L,v_L) + (a^2 - b^2) c_L(v_L,v_L)  \geq \nonumber \\
&&a_L(v_L,v_L) + a b_L(v_L,v_L) + (1- \alpha^2)a^2 c_L(v_L,v_L)  \geq \nonumber \\
&&C (|v_L|_{1,\omega_h}^2 + a^2 ||v_L||_{0,\omega_h}^2)- \alpha^2a^2 c_L(v_L,v_L).
\end{eqnarray}
Thus there exist $\alpha$, $0 < \alpha <1$ and $C>0$ such that if $|a| \geq \beta_0$ and $|b| \leq \alpha |a|$, then for all $v_L \in H^1(\omega_h,\CC^2)$,
\begin{equation}
\label{eq103c}
{\mbox {Re}}\left\{a_L(v_L,v_L) + \beta b_L(v_L,v_L) + \beta^2 c_L(v_L,v_L)\right\} \geq
C ||v_L||_{1,\omega_h}^2.
\end{equation}
Therefore, as in the case where $\beta \in \RR$, with these values of $\beta_0$ and $\alpha$, if $|a| \geq \beta_0$ and $|b| \leq \alpha |a|$, it follows that $i\beta \in \rho({\cal A})$ and $({\cal A} - i \beta I)^{-1}$ is a compact operator in ${\cal H}$.

In order to get more precise estimates of the resolvent, we shall use the theory of elliptic problems with a parameter (\cite{Agranovich}). Let us outline the results of \cite{Agranovich} we shall apply (\cite{Agranovich}, Chapter I).

Let $G$ be an open set of $\RR^n$ ($n \in \NN^*$) with a ${\cal C}^{\infty}$ boundary satisfying hypotheses of section 1.9 of \cite{Agranovich}.
Consider the system of equations
\begin{equation}
\label{eq17c}
A(x,D,q) u(x,q)= f(x,q) \mbox{ in } G,
\end{equation}
where $u$ and $f$ are vector functions with values in $\CC^N$ ($N \in \NN^*$), $A$ is a square matrix of order $N$ consisting of differential operators in $x$ with complex coefficients that have a polynomial dependence on a parameter $q$ and are ${\cal C}^{\infty}$ with respect to $x \in \overline{G}$, $D$ = $(D_1, \ldots, D_n)$, $D_j= -i \partial_{j}$, $j=1, \ldots,n$. The parameter $q$ varies in a sector of the complex plane $ \alpha \leq$ arg$q \leq \beta$ denoted by $Q$.
On the boundary $\partial G$ we are given the conditions:
\begin{equation}
\label{eq17d}
B_j(x,D,q) u(x,q)= g_j(x,q) \mbox{ on } \partial G,\, j=1,\ldots,r.
\end{equation}
Here $B_j$ is a row of order $N$ consisting of differential operators in $x$ with complex coefficients that have a polynomial dependence on the parameter $q$ and are ${\cal C}^{\infty}$ with respect to $x \in \overline{G}$ and $g_j$ is a function with values in $\CC$.
The symbols $A(x,\xi,q)$ and $B_j(x,\xi,q)$ are polynomial in $(\xi,q)$ = $(\xi_1,\ldots,\xi_n,q)$ of degree $s$ and $m_j$. Let us denote by $A_0(x,\xi,q)$ and $B_{j0}(x,\xi,q)$ the principal parts of $A(x,\xi,q)$ and $B_j(x,\xi,q)$ formed of homogeneous polynomials of degree $s$ and $m_j$ in $(\xi,q)$ and set $l_0= max(s,m_1+1, \ldots, m_r+1)$. We now state the two conditions under which estimates of the solutions of \eqref{eq17c}, \eqref{eq17d} can be established.
\begin{condition}
\label{cond1}
If $x \in \overline{G}$, $q \in Q$, $|\xi|+ |q| \neq 0$, then det$A_0(x,\xi,q) \neq 0$. If $n=1$, it is assumed that for $x \in \overline{G}$, $q \in Q$, $q \neq 0$, the roots of the equation in $\lambda$: det$A_0(x,\lambda,q)=0$ are equally distributed between the upper and lower half-plane. The number $r$ of boundary conditions is taken to be $\frac{Ns}{2}$.
\end{condition}
\begin{condition}
\label{cond2}
If $x' \in \partial G$, we suppose that the operators $A$ and $B_j$ are written in the system of coordinates connected with this point (see \cite{Agranovich}, section 1.9). We consider the problem on the half-line (with the notation $D_y= -i \partial_y$ and $\xi'=(\xi_1, \ldots,\xi_{n-1})$)
\begin{equation}
\label{eq17e}
A_0(0,\xi', D_y,q) v(y)= 0, \, y>0,
\end{equation}
\begin{equation}
\label{eq17f}
\{B_{j0}(0,\xi',D_y,q) v\}(y=0)= h_j,\, j=1,\ldots,r.
\end{equation}
It is required that if $|\xi'|+|q| \neq 0$, $q \in Q$, this problem should have for any $h_j$ one and only one solution in the class ${\cal M}$ of stable solutions of \eqref{eq17e}.
\end{condition}
Under conditions \ref{cond1} and \ref{cond2} the following fundamental result holds true (\cite{Agranovich}, Theorems 6.1 and 6.2):
\begin{theoreme}
\label{th1}
Suppose that problem \eqref{eq17c}, \eqref{eq17d} satisfies conditions \ref{cond1} and \ref{cond2} and that $l$ is an integer $\geq l_0$.  Then for $q \in Q$ with sufficiently large moduli, if $f \in H^{l-s}(G, \CC^N)$ and $g_j \in H^{l-m_j-1/2}(\partial G,\CC)$, problem \eqref{eq17c}, \eqref{eq17d} has a unique solution $u \in H^l(G,\CC^N)$. Moreover there exists a constant $C>0$ such that for $q \in Q$ with sufficiently large moduli,
\begin{equation}
\label{eq17g}
|||u|||_{l,G} \leq C(|||f|||_{l-s,G} + \sum_{j=1}^r |||g_j|||_{l-m_j-1/2, \partial G}).
\end{equation}
\end{theoreme}
In \eqref{eq17g}, we have used the following notation: if $m \in \RR$, $m\geq 0$, $|||.|||_{m,G} = (||.||^2_{m,G}+ |q|^{2m} ||.||^2_{0,G})^{1/2}$ (and a similar notation for $|||.|||_{m, \partial G}$). Moreover if $m \in \NN$, due to the interpolational inequality (\cite{Agranovich}, pp.61-62), there exist constants $C_1$ and $C_2 >0$ such that for all $u \in H^m(G,\CC^N)$, 
\begin{equation}
\label{eq17h}
C_1|||u|||_{m,G} \leq (\sum _{k=0}^m |q|^{2k} ||u||^2_{m-k,G})^{1/2} \leq C_2|||u|||_{m,G}.
\end{equation}
If $n=1$, in Theorem \ref{th1}, $g_j$ is assumed to be in $\CC$ and  examining the proof in \cite{Agranovich}, pp.71-72, in \eqref{eq17g} one must replace $\sum_{j=1}^r |||g_j|||_{l-m_j-1/2, \partial G}$ by 
$|q|^{l-m_j-1/2}|g_j|$.

The system \eqref{eq28} may be written under the form:
\begin{equation}
\label{eq17a}
L(D_{1},\beta) U_1 = A D_{11} U_1 + \beta B D_{1} U_1 + (\beta^2 C - \omega^2 \rho) U_1 = i \beta C F_1 + B \partial_1 F_1 + C F_2 \mbox{ in } \omega_h,
\end{equation}
where
\begin{equation}
\label{eq17b}
L(\xi,\beta) = A \xi^2 + \beta B \xi + \beta^2 C - \omega^2 \rho I.
\end{equation}
On the other hand, 
equation \eqref{eq27b} may be written under the form:
\begin{equation}
\label{eq18a}
\{M(D_1, \beta)U_1\}(\pm h) = A D_{1}U_1(\pm h) + \beta D U_1(\pm h) = i D F_1(\pm h), 
\end{equation}
where
\begin{equation}
\label{eq18b}
M(\xi, \beta) = A \xi + \beta D.
\end{equation}
Let $\theta_0 \in \RR$ be such that $0 < \theta_0 < \pi/2$ and suppose that $\beta \in B_{\theta_0}$ where
\begin{equation}
\label{eq108}
B_{\theta_0} = \left \{ \beta = |\beta| e^{i \theta}, \, |\theta| \leq \theta_0 \mbox{ or } |\theta -\pi| \leq \theta_0 \right \}.
\end{equation}
We shall show that the operators $L(D_{1},\beta)$ and $M(D_1, \beta)$ satisfy conditions \ref{cond1} and \ref{cond2} when $\beta \in B_{\theta_0}$.
\begin{lemme}
\label{lem1}
The operator $L(D_{1},\beta)$ satisfies condition \ref{cond1} when $\beta \in B_{\theta_0}$.
\end{lemme}
{\bf Proof} The principal part of $L(\xi,\beta)$ is 
\begin{equation}
\label{eq106}
L_0(\xi,\beta) = A \xi^2 + \beta B \xi +\beta^2 C =
\left(\begin{array}{cc}
(\lambda+2\mu)\xi^2 + \mu \beta^2 & (\lambda+\mu) \xi \beta \\
(\lambda+\mu)\xi \beta & \mu\xi^2 + (\lambda + 2 \mu) \beta^2
\end{array}
\right),
\end{equation}
so that
\begin{equation}
\label{eq107}
\det L_0(\xi,\beta) = 
((\lambda+2\mu)\xi^2 + \mu \beta^2)(\mu\xi^2 + (\lambda + 2 \mu) \beta^2)-
(\lambda+\mu)^2 \xi^2 \beta^2=
(\lambda+2 \mu)\mu (\xi^2 + \beta^2)^2.
\end{equation}
Consequently if $\beta \in B_{\theta_0}$ and $|\xi|+ |\beta| \neq 0$ then $\det L_0(\xi,\beta) \neq 0$. On the other hand for $\beta \in B_{\theta_0}$ and $\beta \neq 0$, the roots of the equation in $\lambda$: det$L_0(\lambda,\beta)=0$ are $\lambda = \pm i \beta$ and thus are equally distributed between the upper and lower half-plane. The number $r$ of boundary conditions is 2 and is equal to $\frac{Ns}{2}$: condition \ref{cond1} is satisfied.
$\Box$
\begin{lemme}
\label{lem2}
The operators $L(D_{1},\beta)$ and $M(D_1, \beta)$ satisfy condition \ref{cond2} when $\beta \in B_{\theta_0}$.
\end{lemme}
{\bf Proof}
For the point of coordinate $x_1 = \gamma h$ ($\gamma = \pm 1$) of the boundary, the interior normal to $\omega_h$ is $- \gamma$. In the neighborhood of the point of the boundary $x_1 = \gamma h$ we choose the local coordinate $y = -\gamma x_1 + h$, so that the boundary point $x_1 = \gamma h$ is such that $y=0$ and the points of the open set $\omega_h$ are locally in the set $(y>0)$.
We must write equation \eqref{eq17a} in a neighborhood of a point of the boundary $x_1 = \gamma h$ in the corresponding local coordinate and take the principal part of the symbol of the corresponding operator, which is $L_0(-\gamma \xi, \beta)$. We must first determine the space ${\cal M}$ of stable solutions of the system:
\begin{equation}
\label{eq18c}
L_0(-\gamma D_y, \beta)w(y)=0 \mbox { in } (y>0).
\end{equation}
The dimension of the space of all the solutions of the system \eqref{eq18c} is 4. Let us first search solutions of \eqref{eq18c} under the form
\begin{equation}
\label{eq109b}
w(y)= W e^{i \alpha y},\,
W= \left( \begin{array}{c} 
W_1 \\
W_3
\end{array}
\right), \,
w= \left( \begin{array}{c} 
w_1 \\
w_3
\end{array}
\right),
\end{equation}
where $\alpha \in \CC$ and $W \neq 0$.
We must have 
\begin{equation}
\label{eq110}
L_0(- \gamma \alpha,\beta) W = 0,
\end{equation}
thus det$L_0(-\gamma \alpha, \beta)=0$ or
\begin{equation}
\label{eq111}
\alpha^2 = - \beta^2 \Leftrightarrow \, \alpha = \eps i \beta,\, \eps = \pm 1.
\end{equation}
Equation \eqref{eq106} implies
\begin{equation}
\label{eq112}
L_0(-\gamma \alpha,\beta) = (\lambda+ \mu) \beta\left(\begin{array}{cc}
-\beta & -\gamma \alpha \\
-\gamma \alpha & \beta
\end{array}
\right)=
(\lambda+ \mu) \beta^2\left(\begin{array}{cc}
-1 & -\gamma \eps i \\
-\gamma \eps i & 1
\end{array}
\right)
\end{equation}
and \eqref{eq110} implies
\begin{equation}
\label{eq113}
-\gamma \alpha W_1 + \beta W_3 =0 \mbox{ or } W_3 = \gamma \eps i W_1.
\end{equation}
Let us search other solutions of \eqref{eq18c} under the form
\begin{equation}
\label{eq114}
w= (X + y Y) e^{i \alpha y} =
(X + y Y) e^{- \eps \beta y},\,
X= \left( \begin{array}{c} 
X_1 \\
X_3
\end{array}
\right), \,
Y= \left( \begin{array}{c} 
Y_1 \\
Y_3
\end{array}
\right), \,
w= \left( \begin{array}{c} 
w_1 \\
w_3
\end{array}
\right)
\end{equation}
where $L_0(-\gamma \alpha,\beta)Y=0$, therefore $Y_3= \gamma \eps i Y_1$.
We have
\begin{equation}
\label{eq115}
D_{y}(y Y e^{i \alpha y})= Y(\alpha y-i) e^{i \alpha y},\,
D_{y}^2(y Y e^{i \alpha y})= Y (\alpha^2 y - 2 i \alpha) e^{i \alpha y},
\end{equation}
thus
\begin{equation}
\label{eq118}
L_0(-\gamma D_y,\beta)(yY e^{i \alpha y}) = [y L_0(-\gamma \alpha, \beta) + L_1(\alpha, \beta, \gamma)]Y e^{i \alpha y},
\end{equation}
with
\begin{equation}
\label{eq119}
L_1(\alpha, \beta, \gamma) =
-i\left(
\begin{array}{cc}
(\lambda+2 \mu) (2 \alpha) & -(\lambda+ \mu) \gamma \beta \\
-(\lambda+\mu) \gamma \beta & \mu(2 \alpha)
\end{array}
\right),
\end{equation}
so that:
\begin{equation}
\label{eq120}
L_1(\eps i \beta, \beta, \gamma)Y =
-i\left(
\begin{array}{cc}
(\lambda+2 \mu) (2 \eps i \beta) & -(\lambda+ \mu) \gamma \beta \\
-(\lambda+\mu) \gamma \beta & \mu(2 \eps i \beta)
\end{array}
\right)
\left(
\begin{array}{c}
1 \\
\gamma \eps i
\end{array}
\right)Y_1=
\eps \beta (\lambda + 3 \mu) \left(
\begin{array}{c}
1 \\
\gamma \eps i
\end{array}
\right)Y_1.
\end{equation}
Equation \eqref{eq18c} where $w$ is given by \eqref{eq114} yields
\begin{equation}
\label{eq123}
[y L_0(-\gamma \eps i \beta, \beta) + L_1(\eps i \beta, \beta, \gamma)] Y + L_0(-\gamma\eps i \beta, \beta) X = 0,
\end{equation}
namely
\begin{equation}
\label{eq124}
\eps \beta (\lambda + 3 \mu) \left(
\begin{array}{c}
1 \\
\gamma \eps i
\end{array}
\right)Y_1
+
(\lambda+ \mu) \beta^2\left(\begin{array}{cc}
-1 & -\gamma\eps i \\
-\gamma \eps i  & 1
\end{array}
\right)
\left(
\begin{array}{c}
X_1 \\
X_2
\end{array}
\right)
= 0,
\end{equation}
or
\begin{equation}
\label{eq125}
\eps \beta (\lambda + 3 \mu) Y_1
+
(\lambda+ \mu) \beta^2
(-X_1 - \gamma \eps i X_2)
= 0.
\end{equation}
Let us choose the following solution of \eqref{eq125}
\begin{equation}
\label{eq126}
Y_1=1, \,X_1 = \eps \frac{\lambda + 3 \mu}{\beta(\lambda + \mu)},\, X_2 =0.
\end{equation}
A basis of solutions of \eqref{eq18c} is formed by $\{w_1^{\eps},\, w_2^{\eps},\, \eps =\pm 1\}$ with

\begin{equation}
\label{eq128}
w_1^{\eps}(y)= 
\left(
\begin{array}{c}
1 \\
\gamma \eps i
\end{array}
\right)
e^{- \eps \beta y},
\end{equation}
\begin{equation}
\label{eq129}
w_2^{\eps}(y)= 
\left[y \left(
\begin{array}{c}
1 \\
\gamma\eps i
\end{array}
\right)+
\left(
\begin{array}{c}
\eps \frac{\lambda + 3 \mu}{\beta(\lambda + \mu)} \\
0
\end{array}
\right)
\right] e^{- \eps \beta y}.
\end{equation}
Therefore we obtain:
\begin{equation}
\label{eq128a}
w_1^{\eps}(0)= 
\left(
\begin{array}{c}
1 \\
\gamma \eps i
\end{array}
\right),
\end{equation}
\begin{equation}
\label{eq129a}
w_2^{\eps}(0)= 
\left(
\begin{array}{c}
\eps \frac{\lambda + 3 \mu}{\beta(\lambda + \mu)} \\
0
\end{array}
\right),
\end{equation}
\begin{equation}
\label{eq130}
\partial_y w_1^{\eps}(y)= 
-\eps \beta \left(
\begin{array}{c}
1 \\
\gamma \eps i
\end{array}
\right)
e^{- \eps \beta y},
\end{equation}
\begin{equation}
\label{eq131}
\partial_y w_2^{\eps}(y)=
\left[(1 - \eps \beta y) \left(
\begin{array}{c}
1 \\
\gamma \eps i
\end{array}
\right) - \eps \beta
\left(
\begin{array}{c}
\eps \frac{\lambda + 3 \mu}{\beta(\lambda + \mu)} \\
0
\end{array}
\right)
\right] e^{- \eps \beta y},
\end{equation}
so that
\begin{equation}
\label{eq130a}
\partial_y w_1^{\eps}(0)=
-\eps \beta \left(
\begin{array}{c}
1 \\
\gamma \eps i
\end{array}
\right)=
 -\left(
\begin{array}{c}
\eps \beta \\
\beta \gamma i
\end{array}
\right)
\end{equation}
and
\begin{equation}
\label{eq131a}
\partial_y w_2^{\eps}(0)=
\left(
\begin{array}{c}
1 \\
\gamma \eps i
\end{array}
\right) -
\left(
\begin{array}{c}
\frac{\lambda + 3 \mu}{\lambda + \mu} \\
0
\end{array}
\right)=
\left(
\begin{array}{c}
-\frac{2 \mu}{\lambda + \mu} \\
\gamma \eps i
\end{array}
\right).
\end{equation}
A basis of the space ${\cal M}$ of stable solutions of \eqref{eq18c} is $\{w_1^{\eps},\, w_2^{\eps}\}$ where $\eps$ is choosen such that Re$(\eps \beta) >0$. Elements of ${\cal M}$ may be written under the form
\begin{equation}
\label{eq127}
w= 
a_1 w_1^{\eps} + a_2 w_2^{\eps}
\end{equation}
where $a_1,a_2 \in \CC$.
The principal part of the symbol of the boundary operator written in the local coordinate $y$ is
$M(-\gamma \xi, \beta)$. 
Now let us establish condition \ref{cond2}.
Given $(h_1,h_2) \in \CC^2$, we must show that there is a unique $w$ of the form \eqref{eq127} satisfying the following system 
\begin{equation}
\label{eq127a}
\{M(-\gamma D_y,\beta)w\}(0)= -A\gamma D_yw(0) + \beta D w(0)= \left(
\begin{array}{cc}
h_1 \\h_2
\end{array}
\right).
\end{equation}
%
But if $w$ is given by \eqref{eq127} then
\begin{equation}
\label{eq127b}
-A\gamma D_yw(0)= 
\left(
\begin{array}{cc}
-(\lambda+2 \mu)i \eps \beta \gamma & - \mu(\lambda+ 2\mu) 2 i \gamma / (\lambda + \mu)\\
\mu \beta & -\eps \mu
\end{array}
\right)
\left(
\begin{array}{c}
a_1\\
a_2
\end{array}
\right)
\end{equation}
and
\begin{equation}
\label{eq127c}
\beta D w(0)= 
\left(
\begin{array}{cc}
\lambda i \eps \beta \gamma & 0\\
\mu \beta & \mu(\lambda+ 3\mu) \eps/ (\lambda + \mu)
\end{array}
\right)
\left(
\begin{array}{c}
a_1\\
a_2
\end{array}
\right).
\end{equation}
The system \eqref{eq127a} boils down to
\begin{equation}
\label{eq127d}
\left(
\begin{array}{cc}
\eps \beta & (\lambda+ 2 \mu)/(\lambda + \mu)\\
\beta & \eps \mu / (\lambda + \mu)
\end{array}
\right)
\left(
\begin{array}{c}
a_1\\
a_2
\end{array}
\right)
=
\frac{1}{2\mu}\left(
\begin{array}{cc}
i \gamma h_1 \\h_2
\end{array}
\right).
\end{equation}
The determinant of this system is $-\beta$ and is $\neq 0$ if $\beta \in B_{\theta_0}$, $\beta \neq 0$: condition \ref{cond2} is satisfied. $\Box$

If  $H$ be a separable Hilbert space and $\{x_n\}_{n=1}^{+\infty}$ is a complete orthonormal sequence in $H$, a bounded linear operator in $H$ is said to be a Hilbert-Schmidt operator if the quantity $||T||_2= (\sum_{n=1}^{+\infty} ||T x_n||^2)^{\frac{1}{2}}$ is finite (\cite{Locker}, p.64).
Theorems \ref{th1} and Lemmas \ref{lem1}, \ref{lem2} imply
\begin{theoreme}
\label{th2}
Let $\theta_0$ be such that $0 < \theta_0 < \pi/2$. Then there exist constants $B>0$ and $C>0$ such that if $\beta \in B_{\theta_0}$ and $|\beta| \geq B$, then $i \beta$ belongs to the resolvent set of ${\cal A}$ and 
\begin{equation}
\label{eq127f}
||({\cal A} -i \beta I)^{-1}|| \leq C.
\end{equation}
Moreover for these values of $\beta$, the resolvent $({\cal A} - i \beta I)^{-1}$ is a Hilbert-Schmidt operator in ${\cal H}$.
\end{theoreme}
{\bf Proof}
Applying Theorem \ref{th1} to equation  \eqref{eq28}(or \eqref{eq17a}) with boundary conditions \eqref{eq27b} (or \eqref{eq18a}), with the values $l=2$, $s=2$, $r=2$, $m_1=m_2=1$, it can be inferred that there exists a constant $B>0$ such that \eqref{eq28}, \eqref{eq27b} has a unique solution for $\beta \in B_{\theta_0}$ and $|\beta| \geq B$ and that there exists a constant $C >0$ such that for all $F = (F_1,F_2) \in {\cal H}$, for these $\beta$,
\begin{eqnarray}
\label{eq127e}
&&||U_1||_{2,\omega_h} + |\beta| ||U_1||_{1,\omega_h} +  |\beta|^2 ||U_1||_{0,\omega_h}
\leq \nonumber \\
&&C(||F_1||_{1,\omega_h} + |\beta| ||F_1||_{0,\omega_h} + ||F_2||_{0,\omega_h}+ |\beta|^{1/2} |F_1(h)|+ |\beta|^{1/2} |F_1(-h)|).
\end{eqnarray}
But by equation $(1.10)$ of \cite{Agranovich} in dimension $n=1$, there exists a constant $C>0$ such that for all $u \in H^1(\omega_h)$, for all $\beta \in \CC$,
\begin{equation}
\label{eq127e1}
|\beta|^{1/2}|u(\pm h)| \leq C(||u||_{1,\omega_h}+ |\beta| ||u||_{0,\omega_h}).
\end{equation}
In view of \eqref{eq127e} and \eqref{eq127e1} we obtain therefrom that there exists a constant $C >0$ such that for all $F = (F_1,F_2) \in {\cal H}$, for all $\beta \in B_{\theta_0}$ such that $|\beta| \geq B$,
\begin{equation}
\label{eq127e2}
||U_1||_{2,\omega_h} + |\beta| ||U_1||_{1,\omega_h} +  |\beta|^2 ||U_1||_{0,\omega_h}
\leq
C(||F_1||_{1,\omega_h} + |\beta| ||F_1||_{0,\omega_h} + ||F_2||_{0,\omega_h}).
\end{equation}
Therefore if $\beta \in B_{\theta_0}$, $|\beta| \geq B$, then $i \beta$ belongs to the resolvent set of ${\cal A}$ and \eqref{eq127e2} and \eqref{eq27} imply that \eqref{eq127f} is satisfied.
Moreover thanks to \eqref{eq127e2} and \eqref{eq27}, for $\beta \in B_{\theta_0}$, $|\beta| \geq B$, $({\cal A} -i \beta I)^{-1}$ is a continuous operator from ${\cal H}= V\times H$ into ${\cal V}= W \times V$.
Since the embeddings from $H^2(\omega_h,\CC^2)$ into $H^1(\omega_h,\CC^2)$ and from $H^1(\omega_h,\CC^2)$ into $L^2(\omega_h,\CC^2)$ are Hilbert-Schmidt operators by Maurin Theorem (\cite{Adams}, p.202), it follows that for $\beta \in B_{\theta_0}$, $|\beta| \geq B$, the resolvent $({\cal A} - i \beta I)^{-1}$ is a Hilbert-Schmidt operator in ${\cal H}$.
$\Box$
\section{Completeness of Lamb modes}
\label{Compl}
\setcounter{equation}{0}
We now summarize some facts about Hilbert spaces.
If $T$ is an unbounded  linear operator in a Hilbert space $H$, let us denote the null space of $T$ by ${\cal N}(T)$ and the range of $T$ by ${\cal R}(T)$. The linear subspaces ${\cal N}(T^n),n \in \NN$, are increasing (${\cal N}(T^n) \subset {\cal N}(T^{n+1}), n \in \NN$) and the linear subspaces ${\cal R}(T^n),n \in \NN$, are decreasing (${\cal R}(T^n) \supset {\cal R}(T^{n+1}), n \in \NN$). The smallest integer $p \in \NN$ such that ${\cal N}(T^p) = {\cal N}(T^{p+1})$ is called the ascent of $T$ and the smallest integer $q \in \NN$ such that ${\cal R}(T^q) = {\cal R}(T^{q+1})$ is called the descent of $T$ (\cite{Locker}, p.30). If $\lambda_0$ is an eigenvalue of $T$ the null space of $T-\lambda_0 I$ is called the eigenspace of $T$ corresponding to $\lambda_0$, the space $\cup_{n=1}^{+\infty} {\cal N}((T-\lambda_0 I)^n)$ is called the generalized  eigenspace of $T$ corresponding to $\lambda_0$ and the elements of this generalized  eigenspace are called generalized eigenvectors (\cite{Locker}, p.33).

A subset of a Hilbert space $H$ is complete (or total) in $H$ if by definition the linear subspace spanned by this set is dense in $H$.
A sequence of vectors $\{u_n\}_{n=1}^{+\infty}$ of a Hilbert space $H$ is a (Schauder) basis if by definition for all $f \in H$ there exists a unique sequence of scalar coefficients $\{c_n(f)\}_{n=1}^{+\infty}$ such that
\begin{equation}
\label{eq130b}
f = \sum_{n=1}^{+ \infty} c_n(f) u_n.
\end{equation}
A sequence of vectors $\{u_n\}_{n=1}^{+\infty}$ of a Hilbert space $H$ (with inner product $(.,.)_H$) is an orthonormal system if by definition $(u_n,u_m)_H = \delta_{nm}$, $n,m =1,\ldots,+\infty$. If a sequence of vectors $\{u_n\}_{n=1}^{+\infty}$ of a Hilbert space $H$ is an orthonormal system then this sequence is a basis iff it is complete (\cite{Christensen_b}, Theorem 3.4.2). But this is not true if the sequence of vectors $\{u_n\}_{n=1}^{+\infty}$ is not an orthonormal system.

If $T$ is an unbounded densely defined closed linear operator in a separable Hilbert space $H$, $T$ is a Hilbert-Schmidt discrete operator iff by definition there exists a point $\alpha \in \rho(T)$ such that the resolvent $(T- \alpha I)^{-1}$ is a Hilbert-Schmidt operator in $H$ (\cite{Locker}, p.78). 

Let us recall some results about Hilbert-Schmidt discrete operators (\cite{Locker}, p.79).
The spectrum of a Hilbert-Schmidt discrete operator in a separable Hilbert space $H$ is a countable set having no finite limit points in $\CC$. Each point $\lambda_0 \in \sigma(T)$ is an eigenvalue of $T$ and the ascent and descent of $T- \lambda_0 I$ are finite and equal (= $p$), the generalized eigenspace ${\cal N}((T-\lambda_0I)^p)$ is finite dimensional with 
\begin{equation}
\label{eq132}
H= {\cal N}((T-\lambda_0 I)^p) \oplus {\cal R}((T-\lambda_0 I)^p)\mbox{ (topological direct sum) }.
\end{equation}
Let $\sigma(T) = \{\lambda_n\}_{n=1}^{+\infty}$ be any enumeration of the spectrum of $T$, let $m_n$ ($m_n \in \NN^*$, $n=1,\ldots,+\infty$) denote the ascent of the operator $T - \lambda_n I$ and let $P_n$, $n=1,\ldots,+\infty$, denote the projection of $H$ onto the generalized eigenspace ${\cal N}((T-\lambda_n I)^{m_n})$ along the range ${\cal R}((T-\lambda_n I)^{m_n})$. Let $S_{\infty}(T)$ be the linear subspace of $H$
\begin{equation}
\label{eq133}
S_{\infty}(T)= \{ u \in H,\, u = \sum_{n=1}^{+\infty} P_n u \}
\end{equation}
and $Sp(T)$ be the linear subspace of $H$ spanned by the set of generalized eigenvectors of $T$. It is easily seen that $\overline{Sp(T)}$ = $\overline{S_{\infty}(T)}$.
 
Let us now state the fundamental theorem of \cite{Locker}, p.80:
\begin{theoreme}
\label{th3}
Let $H$ be a separable Hilbert space, let $T$ be a Hilbert-Schmidt discrete operator in $H$.
Suppose there exists a set of five rays $\gamma_j$: arg $\lambda = \theta_j$, $j =1,\ldots,5,$ in the complex plane such that
 
(i) the angles between adjacent rays are $< \pi/2$,

(ii) for $|\lambda|$ sufficiently large all the points on the five rays belong to $\rho(T)$, and

(iii) there exists $N \in \NN$ such that the resolvent of $T$ satisfies:
\begin{equation}
\label{eq37}
||(T- \lambda I)^{-1}|| = O(|\lambda|^N) \mbox { as } |\lambda| \rightarrow +\infty \mbox{ along each ray } \gamma_j.
\end{equation}
Then $\overline{Sp(T)} = \overline{S_{\infty}(T)} = H$, that is the set of generalized eigenvectors of $T$ is complete in $H$.
\end{theoreme}
Theorem \ref{th2} shows that ${\cal A}$ is a Hilbert-Schmidt discrete operator in ${\cal H}$.

Before stating Theorem \ref{th4}, let us introduce the polynomial operator pencil $P$ associated to Lamb modes. This polynomial operator pencil $P$ is defined by: for all $\mu \in \CC$, for all $v \in H^2(\omega_h,\CC^2)$,
\begin{equation}
\label{eq37a}
P(\mu)v= \left(\begin{array}{c}
P^I(\mu) v\\
P^B(\mu) v
\end{array}
\right),
\end{equation}
with
\begin{equation}
\label{eq37b}
P^I(\mu)v = A \partial_{11}v + \mu B \partial_1 v + \omega^2 \rho v +\mu^2 C v \in L^2(\omega_h,\CC^2)
\end{equation}
and
\begin{equation}
\label{eq37c}
P^B(\mu)v = (A \partial_{1}v(h) + \mu D v (h), A \partial_{1}v(-h) + \mu D v (-h)) \in \CC^2.
\end{equation}
With this definition of $P$, $v_L \in H^2(\omega_h,\CC^2)$, $v_L \neq 0$ is a Lamb mode corresponding to the Lamb eigenvalue $\mu = i \beta$ ($\beta \in \CC$) iff $P(\mu) v_L=0$. If $v_L \in H^2(\omega_h,\CC^2)$, $v_L \neq 0$ is a Lamb mode corresponding to the Lamb eigenvalue $\mu = i \beta$, a family of vectors $v_1, \ldots, v_k$ $\in H^2(\omega_h,\CC^2)$ is said to be associated to the Lamb mode $v_0=v_L$ iff
\begin{equation}
\label{eq37d}
P(\mu)v_p + P'(\mu) v_{p-1}+ (P''(\mu)/2) v_{p-2} + \cdots +(P^{(p)}(\mu)/p!) v_{0} = 0, \, p=0,\ldots,k.
\end{equation}
The vectors $v_1, \ldots, v_k$ are called associated modes.
Applying Theorems \ref{th2} and \ref{th3} we obtain
\begin{theoreme}
\label{th4}
The set of Lamb modes and associated modes is complete in $V= H^1(\omega_h, \CC^2)$.
\end{theoreme}
{\bf Proof} Apply Theorem \ref{th2} with $\theta_0$ such that $2 \pi/5 < \theta_0 < \pi/2$ and Theorem \ref{th3} with $\theta_j = 2(j-1)\pi/5 +\pi/2$, $j=1,\ldots,5$.
Hence the set of generalized eigenvectors of ${\cal A}$ is complete in ${\cal H}$. It is easily seen that this implies the completeness in $V$ of the projection on $V$ of the set of generalized eigenvectors of ${\cal A}$.
If $U =(U_1,U_2)$ $\in D({\cal A})$ is an eigenvector of ${\cal A}$ corresponding to the eigenvalue $i \beta \in \CC$, then \eqref{eq27}, \eqref{eq28}, \eqref{eq27b} are satisfied with $F=0$. In that case $U_1 \in H^2(\omega_h,\CC^2)$ is a Lamb mode corresponding to the Lamb eigenvalue $i\beta$ or with the notations \eqref{eq37a}, \eqref{eq37b}, \eqref{eq37c}, $P(i \beta) U_1=0$.
If $U =(U_1,U_2)$ $\in D({\cal A})$ is a generalized eigenvector of rank $m$ of ${\cal A}$ corresponding to the eigenvalue $i \beta \in \CC$, that is to say $({\cal A} - i \beta I)^m U=0$ and $({\cal A} - i \beta I)^{m-1} U \neq 0$ ($m \in \NN^*$), let $(U^j)_{j=0,\ldots,m}$,  be the chain generated by $U$ defined by $U^m = U$, $U^{j} = ({\cal A} - i \beta I)^{m-j} U^m$, $j=0,\ldots,m$ ($U^{j}$ is a generalized eigenvector of rank $j$ of ${\cal A}$ corresponding to the eigenvalue $i \beta \in \CC$, $j=1,\ldots,m$). With the notations \eqref{eq37a}, \eqref{eq37b}, \eqref{eq37c}, the relation $({\cal A} - i \beta I) U^{j+1} = U^j$, $j=0, \ldots, m-1$ implies $U_1^j \in H^2(\omega_h, \CC^2)$ ($j=1, \ldots,m$), $P(i\beta) U_1^1=0$  and
\begin{equation}
\label{eq38}
P(i\beta) U_1^{j+1}+ P'(i \beta) U_1^{j} + \frac{P''(i \beta)}{2} U_1^{j-1}=0, \, j=1,\ldots, m-1.
\end{equation}
Therefore $U_1^1$ is a Lamb mode corresponding to the Lamb eigenvalue $i \beta \in \CC$ and the vectors $U_1^j$, $j=2,\ldots, m$ (and in particular $U_1= U_1^m$) are associated modes. Consequently we have shown that the set of Lamb modes and associated modes is complete in $V= H^1(\omega_h, \CC^2)$.
$\Box$
\begin{remarque}
\label{rem1}
We have considered the case of a traction-free plate on the upper and lower boundary. But the case of a clamped plate on the upper or lower boundary  can easily be carried out.  Assume for example that the plate is clamped on the lower boundary and traction-free on the upper boundary. Let $V_0$ be the space $V_0= \{v \in H^1(\omega_h,\CC^2), v(-h)=0\}$. In \eqref{eq13b}, the space $H^1(\omega_h,\CC^2)$ (=$V$) must be replaced by $V_0$. Equation \eqref{eq23} must be replaced by $V_L(-h)=0$ and $\tilde{B} V_L(h)=0$. In the proof of Lemma \ref{lem2} \eqref{eq127a} must be replaced by
\begin{equation}
\label{eq127g}
w(0)= \left(
\begin{array}{cc}
h_1 \\h_2
\end{array}
\right).
\end{equation}
But thanks to \eqref{eq128a} and \eqref{eq129a} equation \eqref{eq127g} has only one solution in the class ${\cal M}$ of stable solutions of \eqref{eq18c} so that condition \ref{cond2} is satisfied.
In the proof of Theorem \ref{th4} the space $H^2(\omega_h,\CC^2)$ must be replaced by the space $H^2(\omega_h,\CC^2) \cap V_0$ and we conclude that the set of Lamb modes and associated modes is complete in $V_0$.
\end{remarque}

%

%
%
\end{document}